\documentclass[%
reprint,
 amsmath,amssymb,
 aps, physrev,
]{revtex4-2}

\usepackage[hidelinks]{hyperref}
\usepackage{svg}

\usepackage{graphicx}
\usepackage{dcolumn}
\usepackage{bm}

\usepackage[utf8]{inputenc}
\usepackage[T1]{fontenc}
\usepackage{mathptmx}
\usepackage{etoolbox}

\begin{document}


\title[Theory of Generalized Hertzian Hyperspheres]{Theory of Generalized Hertzian Hyperspheres}
\author{Ulf R. Pedersen}
\email{ulf@urp.dk}
\affiliation{Glass and Time, IMFUFA, Department of Science and Environment, Roskilde University, P.O. Box 260, DK-4000 Roskilde, Denmark}

\date{\today}

\begin{abstract}
While hard-sphere models form the foundation of theoretical condensed matter physics, real systems often exhibit some degree of softness. We present a theoretical and numerical study of a class of nearly hard-sphere systems, generalized Hertzian hyperspheres, where particles interact via a finite-range repulsive potential that allows slight overlaps. Well-studied examples of this class include particles with harmonic repulsions, Hertzian spheres, and Hertzian disks. We derive closed-form expressions for thermodynamic properties, coexistence pressures, and scaling laws governing structure and dynamics. The theory predicts how quantities scale with temperature, density, spatial dimension, and potential softness. These theoretical predictions are tested through numerical simulations in dimensions ranging from one to eight.
\end{abstract}

\maketitle


\section{Introduction}
In this work, we investigate \emph{generalized Hertzian hyperspheres}, constituting a family of \emph{nearly} hard spheres.
Hard spheres play a fundamental role in understanding matter. They represent a simplification of more complex interactions that allow for simpler explanations of fluids and solids' structural and dynamic characteristics. Notable examples are the kinetic theories of gases pioneered by Bernoulli \cite{Bernoulli1738}, Bernal's models for atomic packings of liquids \cite{Bernal1964}, the seminal numerical simulations by Alder and Wainwright demonstrating crystallization \cite{Alder1957}, and the perturbation and integral equations of simple liquids developed in the 1960s \cite{Hansen2013}.
The hard-sphere reference has generally been used to describe atomic, molecular, and colloidal systems.
The packing of spheres is of interest outside of physics as a fundamental problem in mathematics \cite{Toth1942,Hales2005, Viazovska2017} and in computer science related to error corrections \cite{Conway1999, Hamming1950}. In these fields, the interest is not limited to two or three dimensions but extends to hypersphere packing in higher dimensions. This is the direct motivation to consider hyperspheres in higher spatial dimensions, but more importantly, high dimensions serve as exploration and development of the theoretical predictions \cite{Charbonneau2011,  Charbonneau2014, Charbonneau2017, Parisi2020, Adhikari2023}.

Here, we consider \emph{nearly} hard spheres where slight overlaps are allowed. For example, this could be elastic spheres \cite{Hertz1882, landau_lifshitz_elasticity} that may deform upon contact. Specifically, we consider thermal particles with a diameter of one where the pair energy of two overlapping particles is defined as $[1-r]^\alpha$ ($r<1$). Here, $r$ is the distance between particles, and $\alpha$ is a softness parameter. When the temperature is $0<T\ll1$, small overlaps are allowed. We refer to this class of systems as \emph{generalized Hertzian hyperspheres} \cite{Miller2011, Zu2016, Fomin2018, Yao2020, Ryzhov2020, Mandal2020, Guo2021, Tsiok2021, Mandal2021, Gaiduk2022, Pmies2009, Yang2011, Pond2011, Mohanty2014, Ouyang2016, Athanasopoulou2017, Muna2019, MartnMolina2019, deJager2021, Boattini2020} in recognition of the elastic contact theory by Hertz \cite{Hertz1882, landau_lifshitz_elasticity}.

Below, we present closed-form theoretical predictions for thermodynamics, the coexistence line, and scaling laws. The theoretical predictions are in excellent agreement with numerical calculations in 1-8 dimensions and various softness ($\alpha$).

The remainder of the paper is structured as follows. In Sec.\ \ref{sec:Generalized_Hertzian_hyperspheres}, we introduce the class of generalized Hertzian hyperspheres, review relevant realizations investigated in the literature, and describe the numerical simulations conducted. We derive closed-form expressions for thermodynamic properties in Sec.\ \ref{sec:thermodynamics}. Section \ref{sec:freezing_lines} demonstrates how the shape of the coexistence line can be predicted using a hard-sphere mapping. Finally, in Sec.\ \ref{sec:scaling_invariance}, we investigate scaling laws for structure and dynamics based on hard-sphere mappings and isomorph theory.

\section{Generalized Hertzian hyperspheres}\label{sec:Generalized_Hertzian_hyperspheres}
 We define the family of \emph{Generalized Hertzian hypersphere} systems as follows: Consider $N$ particles in $d$ dimensional space. Let ${\bf R} = ({\bf r}_1, {\bf r}_2, \ldots, {\bf r}_N)$ be the collective coordinate vector (with $dN$ elements) of particles confined in a finite volume $V=L^d$ with periodic boundaries (the surface of a flat hypertorus), where $L$ is the length of the sides in a hypercube. The density is $\rho=N/V$. Let the potential energy surface be given by a sum of pair potentials
\begin{equation}
    U({\bf R}) = \sum_{n>m}^N v(r_{nm})
\end{equation}
where $r_{nm}=|{\bf r}_m-{\bf r}_n|$ is the pair distance between particle $n$ and $m$. In this study, we limit ourselves to pair potentials of the form
\begin{equation}\label{eq:pair_potential}
	v(r) = \left[1-r\right]^\alpha \textrm{ for } r\leq1
\end{equation}
and zero otherwise. Here and throughout the paper, we use reduced units, where the particle diameter, the Boltzmann constant and particle masses are all unity.

\begin{figure}
    \centering
    \includegraphics[width=0.8\linewidth]{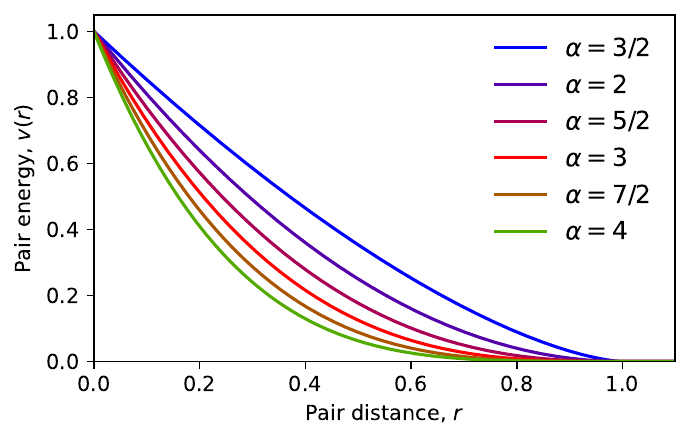}
    \caption{Pair potentials, $v(r)$, of generalized Hertzian hyperspheres, Eq.\ (\ref{eq:pair_potential}), for softness parameters, $\alpha$'s, investigated in this study.}
    \label{fig:pair_potential}
\end{figure}

The generalized Hertzian hyperspheres represent a vast class of systems. A handful of special cases are well-studied: Hertzian spheres\cite{Pmies2009, Yang2011, Pond2011, Mohanty2014, Ouyang2016, Athanasopoulou2017, Muna2019, MartnMolina2019, deJager2021} are given by $d=3$ and $\alpha=\frac{5}{2}$, and Hertzian disks\cite{Miller2011, Zu2016, Fomin2018, Yao2020, Ryzhov2020, Mandal2020, Guo2021, Tsiok2021, Mandal2021, Gaiduk2022} are given by $d=2$ and $\alpha=\frac{7}{2}$. The exponents $\alpha=\frac{5}{2}$ and $\alpha=\frac{7}{2}$ results from the repulsion of two slightly deformed elastic spheres or disks \cite{Athanasopoulou2017, Boattini2020} as famously derived by Hertz \cite{landau_lifshitz_elasticity}. Hertzian spheres are commonly used to model colloidal suspensions \cite{Pmies2009, Charbonneau2011, Charbonneau2017, Scotti2022, Mohanty2014, Royall2024}.

Other well-studied systems are the repulsive harmonic potentials, $\alpha=2$, in two, three or higher dimensions \cite{Jacquin2010, Berthier2010, Zhu2011, Mohanty2014, Levashov2017, Xu2019, MartnMolina2019, Levashov2019, Levashov2020, Santo2021, Levashov2022}. The exponent $\alpha=2$ is often used as a low-temperature approximation for other potentials like the noted Weeks-Chandler-Anderson (WCA) potential \cite{Andersen1971, Valds2018, Nandi2021, Singh2021, Banerjee2021, Zhou2022, Attia2021, Attia2022}. The WCA potential is central in theories of simple liquids. The $\alpha=2$ parametrization is also commonly used in coarse-grained modeling of molecular systems such as in the \emph{dissipative particle dynamics} method \cite{Zhu2011, Levashov2019, Levashov2017, Levashov2020, Santo2021}.
At sufficiently low densities and temperatures, spheres rarely overlap ($r<1$) since Boltzmann factors become large. Thus, in that part of the thermodynamic phase diagram (including a part of the coexistence line), the system approaches that of hard spheres with diameters $\sigma=1$. 

\subsection{Numerical Simulation}
Numerical simulations presented below are conducted using the software packages RUMD \cite{RUMD}, gamdpy \cite{GAMDPY}, and dompap \cite{dompap}, which offer fast computations by being highly parallelized. From an initial state (${\bf R}$ and $\dot{\bf R}$ using Newton's notation for time-derivative), we construct trajectories, ${\bf R}(t)$, within the canonical ensemble (constant $NVT$) by integration of the Langevin force equation:
\begin{equation}
    \ddot{\bf R}(t) = -\nabla U({\bf R}(t))-\xi \dot {\bf R}(t) +\sqrt{2\xi T}{\bf \eta}(t)
\end{equation}
where $-\nabla U({\bf R}(t))$ is a force vector, $\xi$ is a friction coefficient, and ${\bf \eta}(t)$ is a vector of delta-correlated Gaussian white noise where the elements have the property
 \begin{equation}
    \langle \eta_i(t) \rangle = 0, \quad
    \langle \eta_i(t) \eta_j(t') \rangle = \delta_{ij} \delta(t - t').
\end{equation}
The Langevin equation is discretized in time using the Leap-Frog G-JF algorithm \cite{GrønbechJensen2014b}.

To analyze trajectories, we compute structural and dynamical observables besides thermodynamic properties. Let 
\begin{equation}\label{eq:gamma_function}
\Gamma(z)\equiv \int_0^\infty t^{z-1}\exp(-t)dt
\end{equation}
be the gamma function with the properties $\Gamma(1)=1$, $\Gamma(\frac{1}{2})=\sqrt{\pi}$ and $\Gamma(z+1)=z\Gamma(z)$. 
If the probability of finding a particle at distance $r$ of another particle is
$
    h(r) = \frac{1}{N}\sum_{n,m}^N\delta(r-r_{nm})
$
(where $\delta(\ldots)$ is an approximation of Dirac's delta function), then the radial distribution function is defined as
\begin{equation}
    g(r) = \frac{h(r)}{\rho S_dr^
{d-1}}
\end{equation}
where
\begin{equation}\label{eq:S_d}
 S_d = \frac{2\pi^{\frac{d}{2}}}{\Gamma(\frac{d}{2})}
\end{equation}
is the surface area of a $d$-dimensional unit sphere (radius one). Specifically, $S_1=2$, $S_2=2\pi\simeq6.3$, $S_3=4\pi\simeq13$, $S_4=2\pi^2\simeq20$, $S_6=\pi^3\simeq31$, and $S_8=\pi^4/3\simeq32$.

Another structural measure is the static structure factor. If
$
    \rho_{\bf q} = \frac{1}{\sqrt{N}}\sum_n\exp(i{\bf q}\cdot{\bf r}_n)
$
is the collective density field (the Fourier transform of the real-space density, $\rho({\bf r})=\sum_n^N\delta({\bf r}-{\bf r}_n)$) then
the static structure factor is
\begin{equation}
    S({\bf q}) = |\rho({\bf q})|^2.
\end{equation}
For fluids, the structure factor only depends on the length of the wave vector, $q=|{\bf q}|$; thus, all information is contained in $S(q)$.

To investigate dynamics, we compute the mean squared displacement,
\begin{equation}
    R^2(t)=\langle |{\bf r}_n(t')-{\bf r}_n(t'-t)|^2 \rangle.
\end{equation}
The diffusion coefficient is computed from the long-time limit:
\begin{equation}
    D = \lim_{t\to\infty} \frac{R^2(t)}{2dt}
\end{equation}

\begin{figure*}
    \centering
\includegraphics[width=0.40\linewidth]{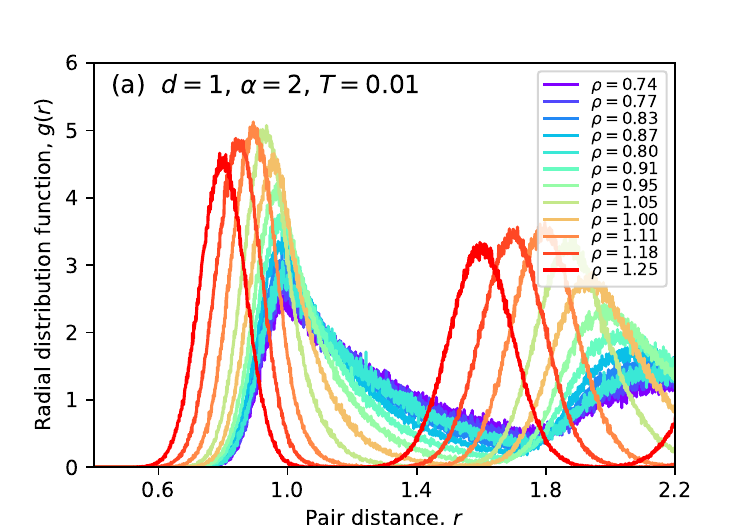}
\includegraphics[width=0.40\linewidth]{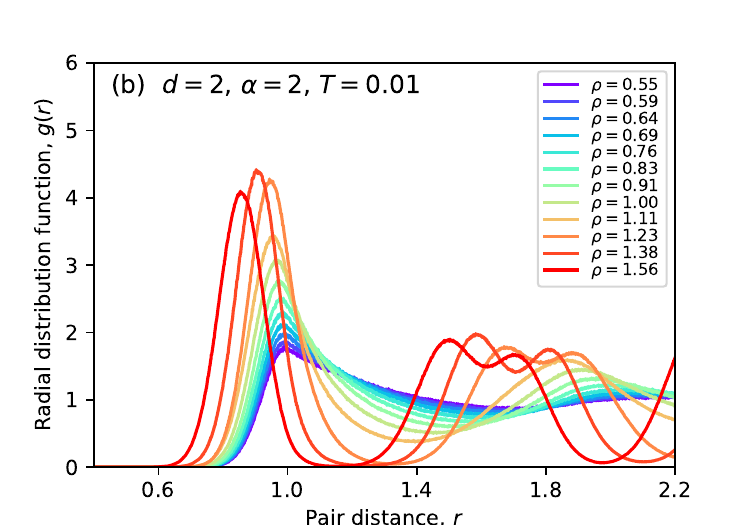} \\
\includegraphics[width=0.40\linewidth]{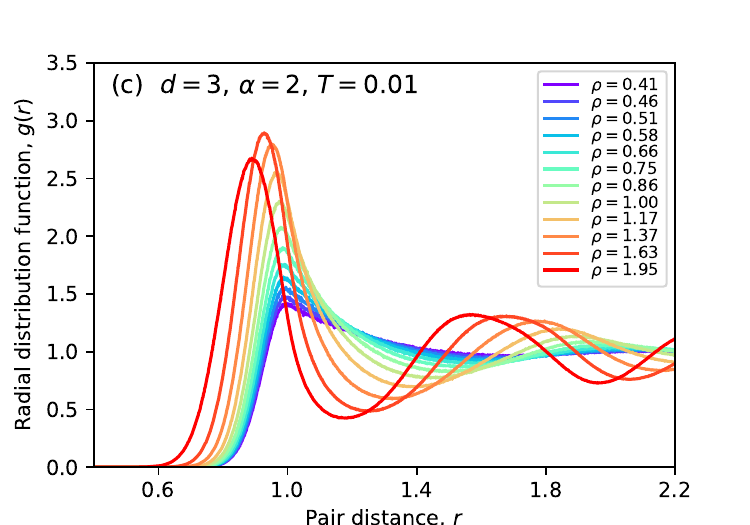}
\includegraphics[width=0.40\linewidth]{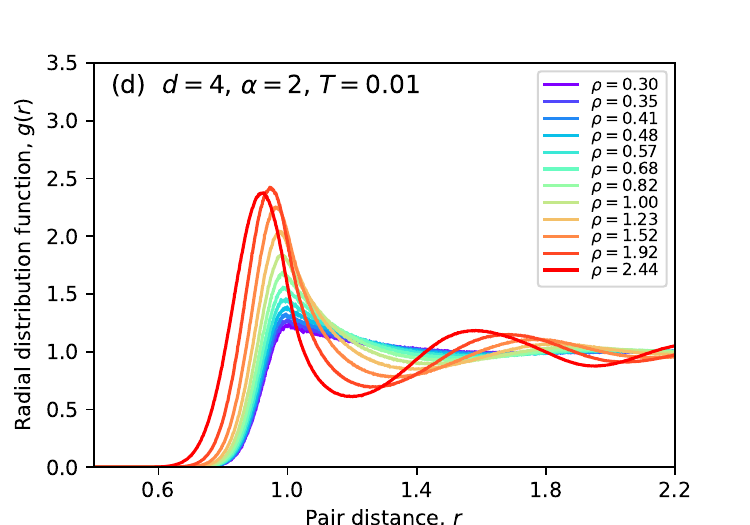} \\
\includegraphics[width=0.40\linewidth]{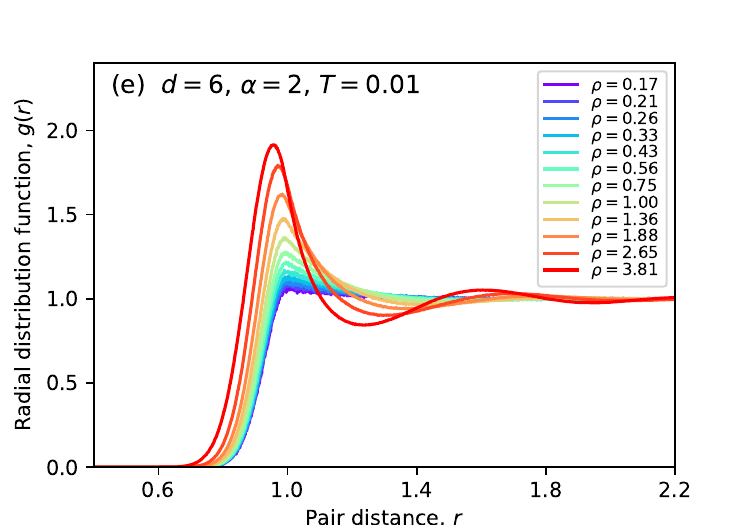}
\includegraphics[width=0.40\linewidth]{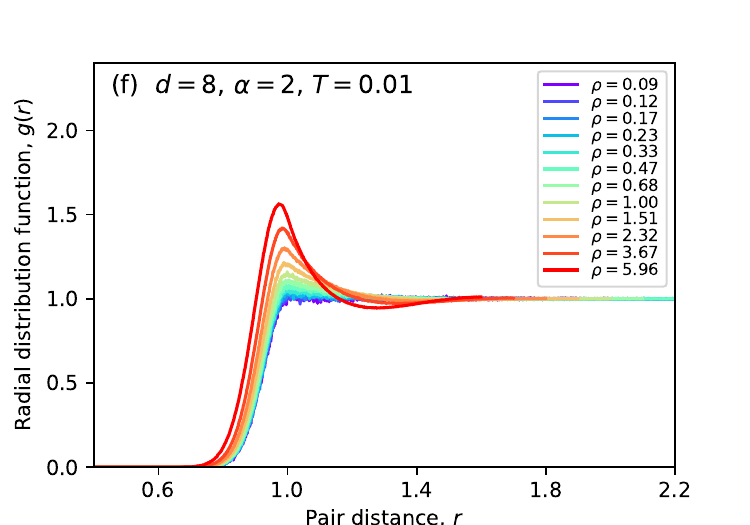} \\
    \caption{Radial distribution function, $g(r)$, for hyperspheres in different dimensions ($d$) with $\alpha=2$ at temperature $T=0.01$. The initial sphere positions are assigned randomly and equilibrated. Each panel depicts $g(r)$ after equilibration.}
    \label{fig:radial}
\end{figure*}

Figures \ref{fig:radial}(a) to \ref{fig:radial}(f) shows $g(r) $'s for representative simulations in 1, 2, 3, 4, 6, and 8 spatial dimensions ($d$), respectively. While simulations in 2 and 3 dimensions are common, simulations in higher dimensions are rare \cite{Lue2006, Lue2010, Charbonneau2011, Bishop2016, Guillard2021, Lue2021, Hoy2022, Hoy2024} (a representative list of references).

\section{Thermodynamics}\label{sec:thermodynamics}
\subsection{Theory}\label{sec:theory}
In general, the partition function for configurational degrees of freedom is given by 
\begin{equation}    
Z_\textrm{conf}=\int_{V^N}\prod_{n=1}^Nd{\bf r}_n \exp\left(-\sum_{n>m}v(r_{nm})/T\right).
\end{equation}
However, when collisions are uncorrelated, we can treat interactions in a mean-field way and write the partition function as 
\begin{equation}
    Z_\textrm{conf}=Z_s^N
\end{equation}
where $Z_s$ is the partition function of a single particle moving in the potential $v_s({\bf r})$ of all other particles frozen in space. 
We will refer to a particle as \emph{free} ($r>1$) if the shorted distance to any other particle is larger than one and \emph{overlapping} ($r<1$) if the distance is shorter than the \emph{kissing} ($r=1$) distance of one. In effect, $Z_s$ has two contributions, one where the moving particle is free, $Z_0$, and one where it overlaps one other particle, $Z_1$:
\begin{equation}\label{eq:Zs_plus_Z1}
    Z_s = Z_0 + Z_1.
\end{equation}
The former is the free volume where a tagged particle never touches another particle. In the low-density limit, where the system approaches an ideal gas, the free volume approaches the entire volume. Thus, the free volume (per particle pair) is
\begin{equation}\label{eq:Z_0_rho_0}
    Z_0\to \rho^{-1}\quad(\rho\to0)
\end{equation}
Another approximation can be achieved by subtracting the volume excluded by other particles.
\begin{equation}\label{eq:Z0_approximation}
    Z_0 = \rho^{-1}-V_d
\end{equation}
where $V_d$ is the volume of $d$ dimensional hypersphere with diameter 1:
\begin{equation}
    V_d = \frac{\pi^\frac{d}{2}}{2^d\Gamma(\frac{d}{2}+1)}.
\end{equation}
Specifically $V_1=1$, $V_2=\frac{\pi}{4}$, $V_3=\frac{\pi}{6}$, $V_4=\frac{\pi^2}{32}$, and so on.
When the density is at or slightly lower than a given close-packed structure, the free volume for the tagged particle will be that of its local cage. Thus, we suggest it can be approximated with
\begin{equation}\label{eq:Z0_approximation_cp}
    Z_0 \simeq \rho^{-1}-\rho_{cp}^{-1}
\end{equation}
where $V_{cp}=N\rho_{cp}^{-1}$ is the volume of the close-packed structure of relevance. (For reference, we note often the \emph{packing fraction}
\begin{equation}
    \eta\equiv\frac{NV_d}{V}
\end{equation}
is sometimes used to give the \emph{compactness} of the system. Then $\rho_{cp}=\frac{\eta_{cp}}{V_d}$ where $\eta_{cp}$ is the packing fraction of the close-packed structure).
For rods on a line, the close packing density ($\rho_{cp}$) is $\rho_{rods}=1$ ($d=1$).
In the plane ($d=2$), hexagonal packing of disk, with $\rho_\textrm{hex}=\frac{2}{\sqrt{3}}\simeq 1.16$, gives the close packing density \cite{Toth1942}. 
The crystalline structures in three dimensions (the fcc, hcp, and related lattices) \cite{Hales2005} have a close-packed density of $\rho_\textrm{fcc}=\sqrt{2}\simeq1.41$, while so-called \emph{random close packing} \cite{Bernal1964} is $\rho_\textrm{rcp}=1.22$ ($d=3$). 
In four-dimensional space ($d=4$) the D4 lattice is the densest known packing \cite{Hoy2024, Conway1999} with a density of $\rho_\textrm{D4}=2$.  
The E6 lattice is the densest known packing in six spatial dimensions ($d=6$). For an eight-dimensional space ($d=8$), M. Viazovska has recently provided a rigorous proof \cite{Viazovska2017} settling that the E8 lattice, with $\rho_\textrm{E8}=16$, yield the densest possible packing of hard-hyperspheres.

The latter term in single particle partition function (Eq.\ (\ref{eq:Zs_plus_Z1})) is taking into account when particles overlap ($r<1$):
\begin{equation}
    Z_1 = \int_0^1 e(r) d{\bf r}
\end{equation}
where
\begin{equation}\label{eq:pair_boltzmann}
	e(r)\equiv \exp(-v(r)/T)
\end{equation}
is the pair-potential Boltzmann probability factor,
\begin{equation}
    d{\bf r} = S_d r^{d-1}  dr
\end{equation}
is the volume element of a spherical integration.

In the low-density limit the expectation value of pairwise quantity $A(r)$, which is zero for $r>1$, is given by \cite{Attia2021}
\begin{equation}\label{eq:general_expectation_value}
    \langle A\rangle = \frac{1}{2Z_s} \int_0^1 A(r) e(r)d{\bf r}.
\end{equation}
where the factor $\frac{1}{2}$ accounts for double counting of pairs. As an example,
$A(r)$ can be considered as the pairwise energy $v(r)$ (Eq.\ (\ref{eq:pair_potential})). Then, the above integral would give the expectation value of the potential energy per particle, 
\begin{equation}
    \langle v\rangle=\frac{U}{N},
\end{equation}
where $U$ is the expectation value of the system's potential energy.
Another option is to let $A(r)$ be the pair virial defined as \begin{equation}
    w(r)\equiv -\frac{r}{d}\frac{dv(r)}{dr}
\end{equation}
that for the generalized Hertzian hypersphere pair potential is
\begin{equation}
    w(r) = \frac{\alpha r}{d}(1-r)^{\alpha-1} \textrm{ for } r\leq1
\end{equation}
and zero otherwise. The above integral would yield \begin{equation}
    \langle w\rangle=\frac{W}{N}
\end{equation} where $W$ is the viral gives the configurational contribution to the pressure $p$,
\begin{equation}\label{eq:pressure}
    p = \rho T + \langle w\rangle.
\end{equation}
Moreover, Eq.\ (\ref{eq:general_expectation_value}) can be used to evaluate the thermodynamic response functions. As an example, consider the specific isochoric heat capacity,
\begin{equation}
    c_V \equiv \frac{1}{N} \left(\frac{\partial E}{\partial T}\right)_V
\end{equation}
where $E=\langle U\rangle+\langle K\rangle$, $\langle U\rangle=N\langle v\rangle$, $K=\sum_n^N\frac{1}{2}|\dot {\bf r}_n|^2$ is kinetic energy and $|\dot {\bf r}_n|$ is the speed of particle $n$ drawn from the Maxwell distribution.
In the canonical ensemble
\begin{equation}
    c_V = \frac{\sigma_v^2}{T^2}+\frac{d}{2}
\end{equation}
where 
\begin{equation}  
\sigma_v^2 \equiv \langle v^2 \rangle - \langle v \rangle^2
\end{equation}
is the variance of the energy fluctuations. Inspired by Refs.\ \cite{Maimbourg2016, Bacher2018, Bacher2018b, Maimbourg2020, Attia2021}, for simplicity, one may consider the low-density limit, $\rho\to0$, where $Z_s\rightarrow0$. We can neglect terms that involve the multiplication of expectation values since terms involving a single expectation value scale are $1/Z_s$. In contrast, terms that involve multiplying expectation values scale as $1/Z_s^2$. Thus, $\sigma_v^2\to\langle v^2\rangle$ for $\rho\to0$, and the isochoric heat capacity can be simply evaluated as
\begin{equation}
    c_V = \frac{\langle v^2\rangle}{T^2}+\frac{d}{2}\quad(\rho\to0)
\end{equation}
To evaluate the thermal pressure coefficient,
\begin{equation}
    \beta_V \equiv \left( \frac{\partial p}{\partial T} \right)_V,
\end{equation}
we need the pressure ($p$; Eq.\ (\ref{eq:pressure})) that involve $\langle w\rangle$.
Thus, to assess thermodynamic properties in general, we need to evaluate
\begin{equation}\label{eq:w_iu_i}
     \langle w^iv^j\rangle=\frac{I_{d\alpha ij}}{2Z_s},
\end{equation}
 where
\begin{equation}\label{eq:Idaij_definition}
    I_{d\alpha ij} \equiv \int_0^1 [w(r)]^i[v(r)]^j e(r) d{\bf r}.
\end{equation}
For $T\to0$ this integral simplifies to (Appendix \ref{appendix:deviation})
\begin{equation}\label{eq:I_daij}
 I_{d\alpha ij}= S_dd^{-i}\alpha^{i-1}T^{J_{\alpha ij}}\Gamma\left(J_{\alpha ij}\right)\quad(T\to0)
\end{equation}
where
\begin{equation}
J_{\alpha ij} \equiv i+j+\frac{1-i}{\alpha}.
\end{equation}
As an example of usage, the \emph{overlap} part of the partition function ($Z_1$ in $Z_s=Z_0+Z_1$) can be written as $Z_1 = I_{d\alpha00}$, or
\begin{equation}\label{eq:Z_1}
    Z_1 = S_dT^{\frac{1}{\alpha}}\Gamma\left(\frac{1}{\alpha}\right)\quad(T\to0)
\end{equation}
The expectation value of energy,
\begin{equation}
 \langle v\rangle =  \frac{I_{d\alpha 01}}{2Z_s}, 
\end{equation}
evaluates to
\begin{equation}\label{eq:theory_v}
    \langle v\rangle = \frac{S_dT^{1+\frac{1}{\alpha}}\Gamma\left(\frac{1}{\alpha}\right)}{2\alpha^2 Z_s}.
\end{equation}
Similar expressions are straightforward to evaluate for quantities such as $\langle u^2\rangle$, $\langle w\rangle$, $\langle w^2\rangle$, or $\langle wu\rangle$.

In the following section, we compare theory with numerical simulations. One caveat of the present theory for thermodynamics is that the approximations for $Z_s$ ignore many-body correlations in particle positions. While crude, they are valuable in explaining how energy scales with temperature and density in particular parts of the phase diagram (see below). For a refined theory for $Z_s$, we suggest to apply ideas from Scaled Particle Theory, Density Functional Theory, or the Percus-Yevick approximation \cite{Wertheim1963, Robles2004, AddaBedia2008, Santos2020} of the Ornstein-Zernike equation \cite{Ornstein1914, Tsednee2019}. We leave such investigations to future studies.

\subsection{Two rods with harmonic repulsion}

\begin{figure}
    \centering
\includegraphics[width=0.8\linewidth]{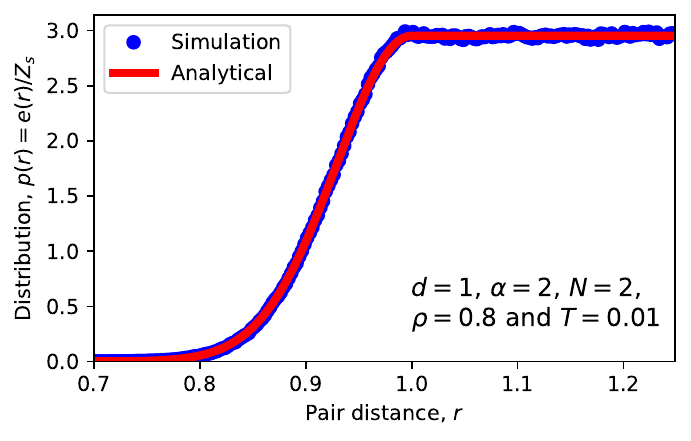}
    \caption{Distribution of pair distance of two harmonic repulsive rods ($\alpha=2$; $d=2$) at temperature $T=0.01$ and density $\rho=0.8$.\label{fig:two_rods_pair_distance}}
    \label{fig:enter-label}
\end{figure}

First, we investigate harmonic repulsive rods with $d=1$ and $\alpha=2$. For the sake of simplicity, consider only two rods ($N=2$) at temperature $T=0.01$ and $\rho=0.8$. Figure \ref{fig:two_rods_pair_distance} shows the pair distance in a numerical simulation of the canonical ensemble. The volume of this system is $V=2.5$, and from both Eqs.\ (\ref{eq:Z0_approximation}) and (\ref{eq:Z0_approximation_cp}) we get $Z_0=0.5$. For this trivial case, Eqs.\ (\ref{eq:Z0_approximation}) and (\ref{eq:Z0_approximation_cp}) are not approximations, and $Z_1=2\sqrt{\pi T}$. 
The probability distribution of the pair distance is 
$
h(r)\equiv e(r)r^{d-1}S_d/Z_s
$
that for this system is
\begin{equation}\label{eq:two_rods_hr}
    h(r) =\frac{\exp\left(-[1-r]^2/T\right)}{\rho^{-1}-1+\frac{1}{2}\sqrt{\pi T}} \quad\text{ for } r<1
\end{equation}
and $1/Z_s$ for $r>1$. Figure \ref{fig:two_rods_pair_distance}(b) compares $h(r)$ of numerical simulation with the analytical Eq.\ (\ref{eq:two_rods_hr}).
The expected value of energy,
\begin{equation}
    \langle u\rangle = \frac{\sqrt{\pi}T^\frac{3}{2}}{\rho^{-1}-1+\frac{1}{2}\sqrt{\pi T}},
\end{equation}
and other statistical properties are trivial in this simple case.

\begin{figure}
    \centering
    \includegraphics[width=0.8\linewidth]{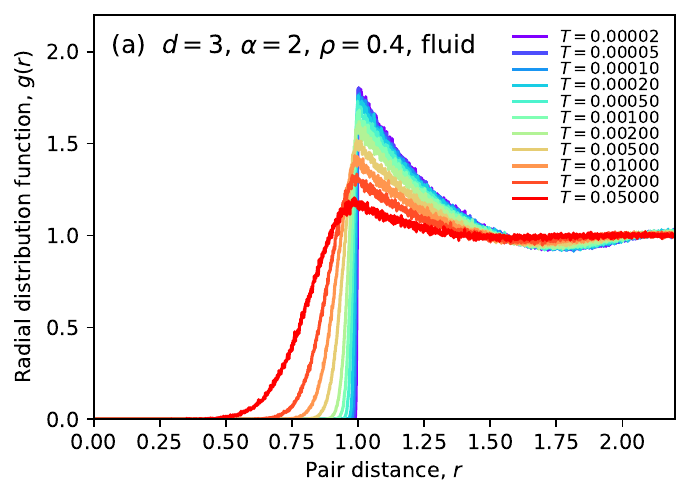}
    \includegraphics[width=0.8\linewidth]{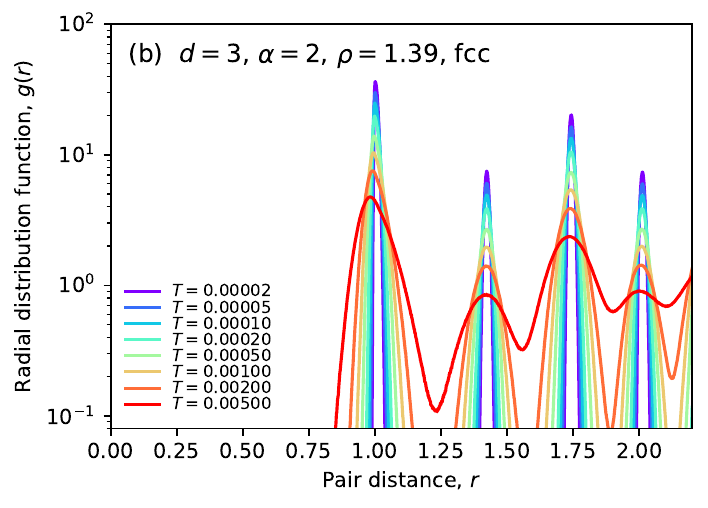}
    \includegraphics[width=0.8\linewidth]{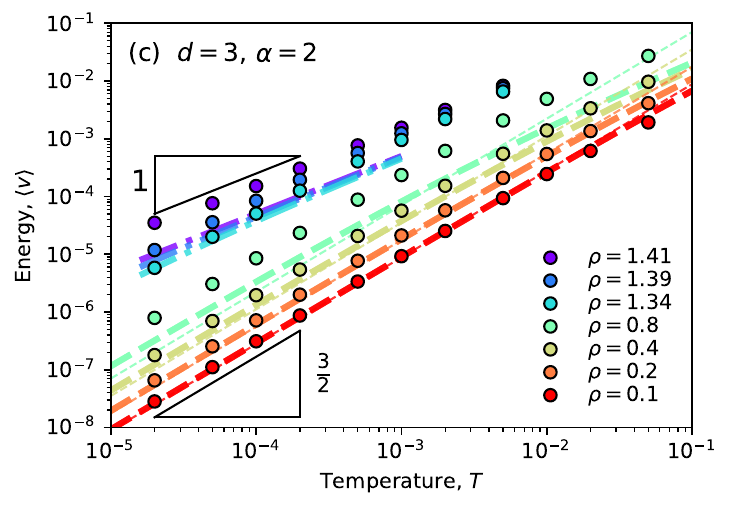}
    \caption{(a) Radial distrubution functions, $g(r)$, for spheres ($d=3$) with harmonic repulsions ($\alpha=2$) in fluid state-points computed in numerical simulations. (b) Radial distribution function, $g(r)$, for state points where particles are in a close-packed fcc crystal. (c) The dots show energies, $\langle v\rangle$ as a function of temperature, $T$, computed in numerical simulations. For the densities $\rho=0.1$ to $\rho=0.8$, the spheres are in a fluid state, and the remaining densities are close-packed fcc crystals. The lines are theoretical prediction (Eq.\ (\ref{eq:theory_v})) using $Z_s=Z_1+\rho^{-1}-V_d$ (thick dashed), $Z_s=\rho^{-1}$ (thin dashed) and $Z_s=Z_1+\rho^{-1}-\rho_\textrm{fcc}^{-1}$ (dash dot).}
    \label{fig:energies_d3}
\end{figure}

\begin{figure}
    \centering
    \includegraphics[width=0.8\linewidth]{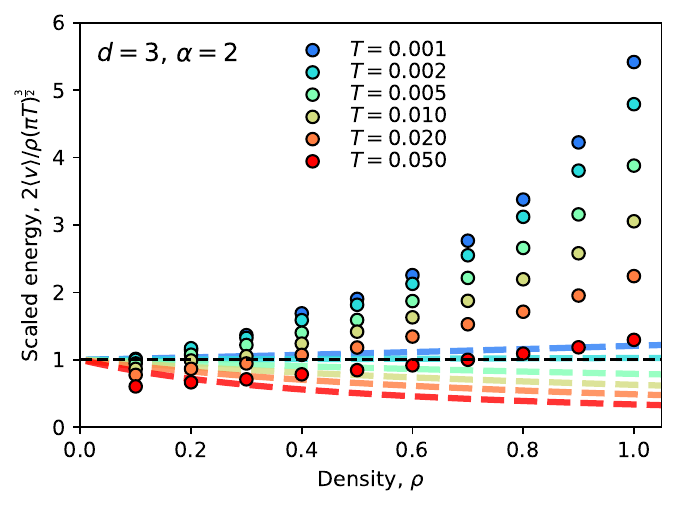}
    \caption{The dots show the scaled energy of harmonic repulsive ($\alpha=2$) spheres ($d=3$) for temperatures $T=(0.001,0.005,0.01)$ as a function density. Energies are scaled with the theoretical prediction of Eq.\ (\ref{eq:theory_v_alpha_2_d3}) with $Z_s=\rho^{-1}$. The dashed lines are the theoretical prediction of Eq.\ (\ref{eq:theory_v_alpha_2_d3}) with  $Z_s=Z_1+\rho^{-1}-V_d$.}
    \label{fig:energy_along_density}
\end{figure}

\subsection{Hyperspheres with harmonic repulsions}\label{sec:harmonic}
Thermodynamics are not trivial for the general case of hyperspheres with harmonic repulsions ($\alpha=2$). Figure \ref{fig:radial} shows the radial distribution function, $g(r)$, for hyperspheres with harmonic repulsions in 1, 2, 3, 4, 6, and 8 dimensions ($d$). 
To compare results for different densities and spatial dimensions, we introduce the characteristic length
\begin{equation}
    \lambda = \rho^{-1/d}.
\end{equation}
Figure \ref{fig:radial} presents results for the temperature $T=0.01$, and densities corresponding to $\lambda$ ranging from 0.8 to 1.25. At low densities, the radial is similar to the Heaviside step function $H(r-1)$, while medium-range structures appear as peaks in $g(r)$ when density increases. This structure is more pronounced at low spatial dimensions than at high.

The potential energy per particle (Eq.\ (\ref{eq:theory_v})) for harmonic repulsions ($\alpha=2$) simplifies to
\begin{equation}\label{eq:theory_v_alpha_2}
    \langle v \rangle = \frac{S_d\sqrt{\pi}}{8Z_s}T^\frac{3}{2}\quad(\alpha=2)
\end{equation}
and
\begin{equation}\label{eq:theory_v_alpha_2_d3}
    \langle v \rangle = \frac{(\pi T)^\frac{3}{2}}{2Z_s} \quad(\alpha=2;\, d=3)
\end{equation}
in three dimensions ($d=3$).
To investigate the validity of this expression, we conducted simulations of spheres ($d=3$) in the fluid state (Fig.\ \ref{fig:energies_d3}(a)), and for the fcc crystal (Fig.\ \ref{fig:energies_d3}(b)) in a range of temperatures from $T=10^{-5}$ to $10^{-1}$. For low temperature and low densities (the fluid), the energy, Fig.\ \ref{fig:energies_d3}(c)), scales as $T^\frac{2}{3}$. This scaling appears as a consequence of $Z_s=\rho^{-1}$ (thin dashed on Fig.\ \ref{fig:energies_d3}(c)). At intermediate densities, say $\rho=0.8$ (teal), however, $Z_s=Z_1+\rho^{-1}-V_d$ gives a better prediction since there is a significant excluded volume. At high densities, in the crystal ($\rho=1.41$), the energy scales as $T$. To explain this, we suggest to use $Z_s=Z_1+\rho^{-1}-\rho_{fcc}^{-1}$ (the dash-dot lines on Fig.\ \ref{fig:energies_d3}(c)).
If $\rho\simeq\rho_{fcc}$ then this simplifies to $Z_s\simeq Z_1$. 
Since $Z_1$ scales as $T^\frac{1}{\alpha}$ (Eq.\ (\ref{eq:Z_1})), 
then Eq.\ (\ref{eq:theory_v_alpha_2}) dictates that $\langle v\rangle$ scales as $T$.

Many-body correlations can be neglected at low densities, making the theoretical prediction more accurate. For example, Fig.\ \ref{fig:energy_along_density} presents the energy of harmoic repulsive spheres ($\alpha=2$, $d=3$) scaled by the theoretical prediction of Eq.\ (\ref{eq:theory_v_alpha_2_d3} with $Z_s=\rho^{-1}$.

\subsection{The role of spatial dimension}

\begin{figure}
    \centering
    \includegraphics[width=0.8\linewidth]{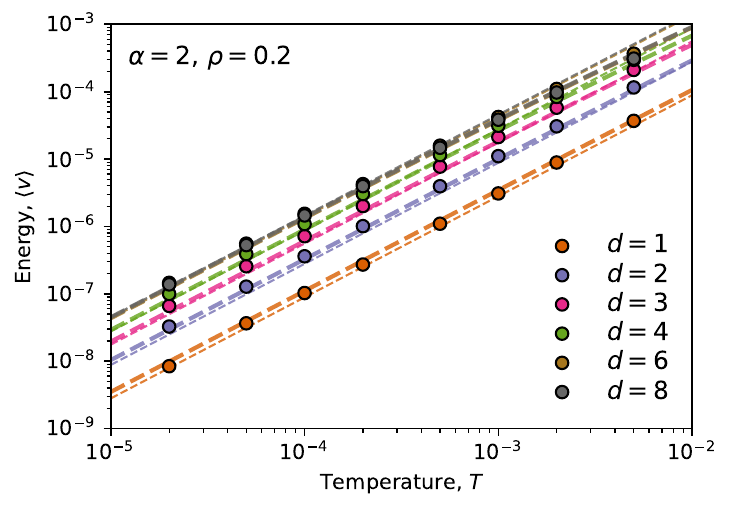}
    \caption{Energies versus temperature for harmonic repulsive ($\alpha=2$) hyperspheres at density $\rho=0.2$. The dots are from numerical simulations. The lines are theoretical prediction (Eq.\ (\ref{eq:theory_v})) using $Z_s=Z_1+\rho^{-1}-V_d$ (thick dashed), and $Z_s=\rho^{-1}$ (thin dashed).}
    \label{fig:energies_vs_density}
\end{figure}

In Eq.\ (\ref{eq:theory_v}) for the energy suggests that for a given temperature ($T$), density ($\rho$), and softness ($\alpha$), the energy should be proportional to the area of a unit sphere, $S_d$ (Eq. (\ref{eq:S_d})). As depicted in Fig.\ \ref{fig:energies_vs_density} the theoretical prediction agrees with numerical simulations.

\begin{figure*}
    \centering
    \includegraphics[width=0.40\linewidth]{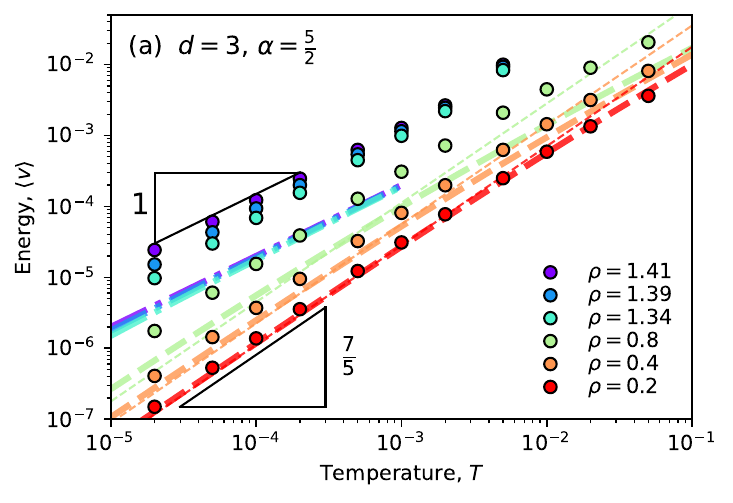}
    \includegraphics[width=0.40\linewidth]{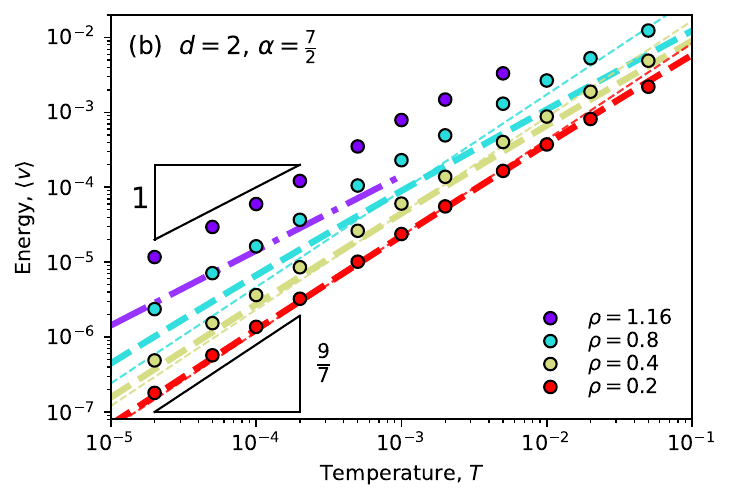}
    \caption{(a) Dots show energy computed by numerical simulations of Hertzian spheres ($d=3$, $\alpha=\frac{5}{2}$) and (b) of Hertzian disks ($d=2$, $\alpha=\frac{7}{2}$). The lines are theoretical prediction (Eq.\ (\ref{eq:theory_v})) using $Z_s=Z_1+\rho^{-1}-V_d$ (thick dashed), $Z_s=\rho^{-1}$ (thin dashed) and $Z_s=Z_1+\rho^{-1}-\rho_\textrm{fcc}^{-1}$ (dash dot). }
    \label{fig:energy_hertzian_spheres}
\end{figure*}

\begin{figure}
    \centering
    \includegraphics[width=0.8\linewidth]{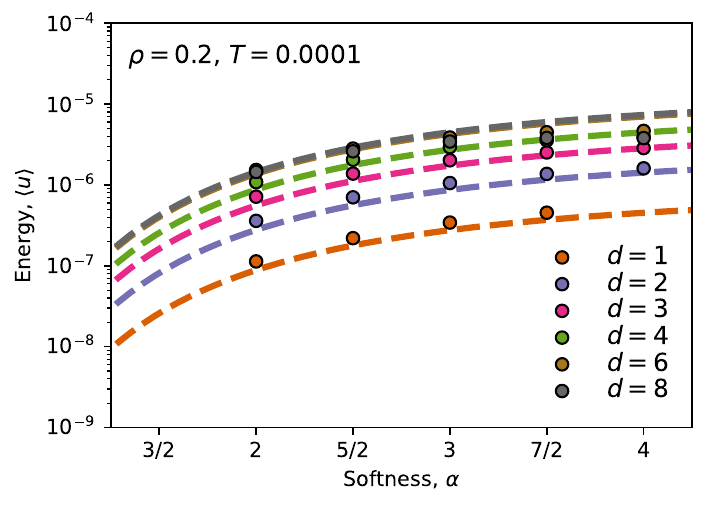}
    \caption{Energy as a function of softness in $d=1$ to $d=8$ at density $\rho=0.2$ and $T=0.0001$.}
    \label{fig:energies_vs_alpha}
\end{figure}

\subsection{The role of softness}

In the previous Section \ref{sec:harmonic}, we investigated the energy particle systems with harmonic repulsions, i.e., with a softness of $\alpha=2$. As suggested by Eq.\ (\ref{eq:theory_v}), the temperature scaling of energy is expected to depend on $\alpha$. As an example, we first investigate the classic Hertzian disks and spheres:

The generalization of the scaling of the energies found for particles with harmonic repulsions predicts that the scaling for fluids at low densities should be $T^{1+\frac{1}{\alpha}}$ and $T$ for solids near the close packing density. For Hertzian spheres ($d=3$, $\alpha=\frac{5}{2}$), the production for the scaling in the fluid evaluates is $T^\frac{7}{5}$, in good agreement with numerical simulations (Fig.\ \ref{fig:energy_hertzian_spheres}(a)). For Hertzian disks ($d=2$, $\alpha=\frac{7}{2}$) scaling in the fluid is $T^\frac{9}{7}$, in good agreement with numerical simulations (Fig.\ \ref{fig:energy_hertzian_spheres}(b)). For both Hertzian disks and spheres, the energy of the solid is proportional to temperature, $T$ (Figs.\ \ref{fig:energy_hertzian_spheres}(a) and \ref{fig:energy_hertzian_spheres}(b)).

Figure \ref{fig:energies_vs_alpha} shows how the energy depends on softness ($\alpha$) in $d=1$ to $d=8$ dimensions for fluids at density $\rho=0.2$ and temperature $T=0.0001$. From the theory it is that energy scales as $T^{1+\frac{1}{\alpha}}$ (dashed lines on Fig.\ \ref{fig:energies_vs_alpha}; Eq. (\ref{eq:theory_v}) with $Z_s=\rho^{-1}$) .

\begin{figure}
    \centering
    \includegraphics[width=0.4\textwidth]{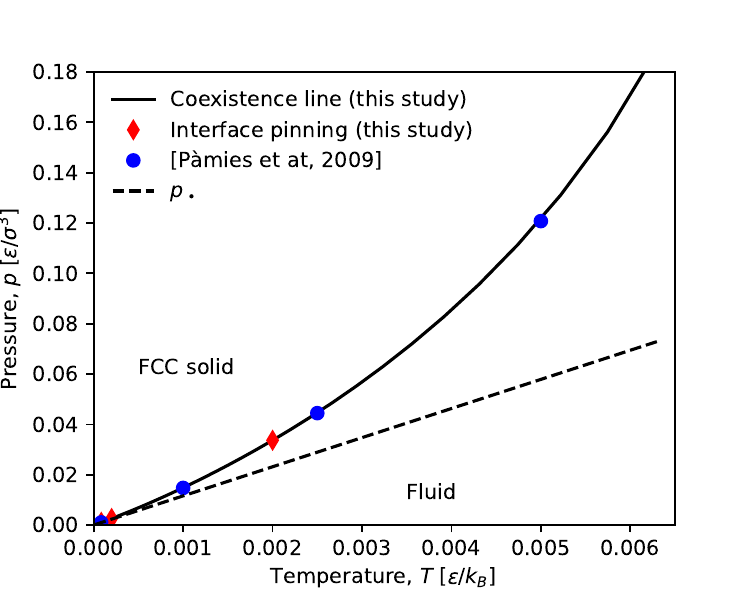}
    \caption{Coexistence pressure of Hertzian spheres ($d=3$; $\alpha=\frac{5}{2}$). The red diamonds are coexistence points computed with the interface pinning method (this study). The solid black line is computed with numerical integration of the Clausius-Clapeyron identity (this study) using coexistence obtained by interface pinning at $T=0.002$. The dashed line is the prediction ($p_\bullet$) of zeroth order hard-sphere mapping, Eq.\ (\ref{eq:p_bullet_hs}).
    }
    \label{fig:pressure}
\end{figure}

\begin{figure}
    \centering
    \includegraphics[width=0.4\textwidth]{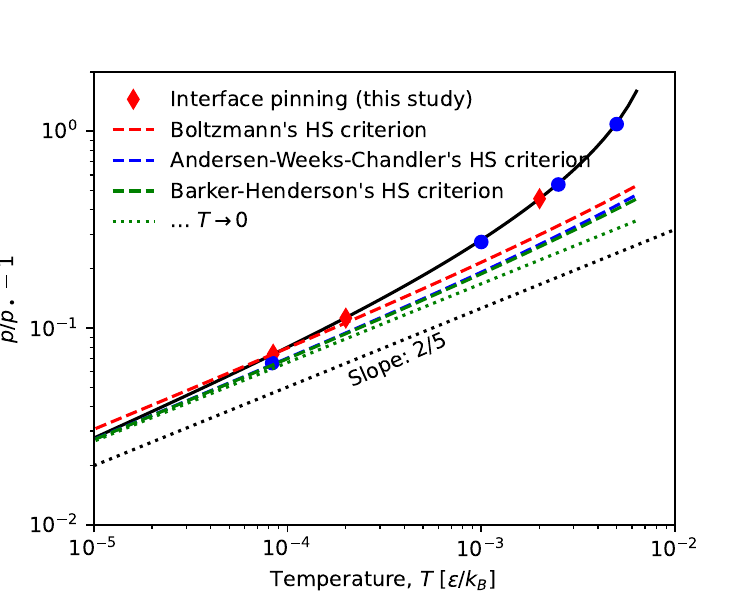}
    \caption{Normalized deviation from zeroth order hard-sphere prediction of coexistence pressure (Fig.\ \ref{fig:pressure}), $\frac{p}{p\bullet}-1$ of Hertzian spheres ($d=3$, $\alpha=\frac{5}{2}$). The solid line, blue dots, and red diamonds are the results of numerical simulations (see Fig.\ \ref{fig:pressure})). The red dashed line is the prediction of the Boltzmann criterion (Eq.\ (\ref{eq:p_boltzmann})), the blue dashed line is from Andersen-Weeks-Chandler theory (Eq.\ \ref{eq:AWC}), and the green dashed line is from Barker-Henderson theory (Eq.\ \ref{eq:p_Barker-Henderson}).}
    \label{fig:pressure_log}
\end{figure}

\section{Freezing lines}\label{sec:freezing_lines}
This section applies the theory of generalized Herzian hyperspheres to make predictions of coexistence-pressure between a solid and a fluid.

\subsection{Zeroth-order hard-sphere prediction}

At sufficiently low densities and temperatures, generalized Herzian hyperspheres approach hard spheres. The most straightforward hard-sphere mapping is to choose the truncation distance as the hard-sphere diameter, that is, setting the hard-sphere diameter to $\sigma=1$.

For example, consider close packing of hard spheres ($d=3$). The coexistence pressure of the hard-sphere system is accurately estimated by Fernandez \textit{et al.} \cite{Fernandez2012} to %
\begin{equation}\label{eq:hs_pressure}
  p_\sigma = 11.5712(10) T/\sigma^3\,.
\end{equation}
When $T\rightarrow0$ for generalized Hertzian spheres, we expect that the coexistence pressure approaches
\begin{equation}\label{eq:p_bullet_hs}
  p_\bullet = 11.5712(10) T\,,
\end{equation}
Throughout this paper, the bullet subscript "$\bullet$" refers to the limit that the potentials approach when $T\rightarrow0$, i.e., setting $\sigma=1$.

\subsection{Boltzmann's hard-sphere criterion}\label{sec:boltzmann}
At any finite temperature, the effective hard-sphere diameter is expected to be less than one since particles are allowed to overlap slightly.
Boltzmann's hard-sphere criterion \cite{Boltzmann1896, Attia2022} is defined as the distance where the pair potential is equal to the thermal energy:
\begin{eqnarray}
    v(\sigma) = T
\end{eqnarray}
that evaluates to
\begin{eqnarray}\label{eq:boltzmann_sigma}
    \sigma = 1-T^\frac{1}{\alpha}.
\end{eqnarray}
for generalized Hertzian hyperspheres. 
Since, $p=p_\bullet/\sigma^d$
then coexistence pressure is given by
\begin{equation}\label{eq:p_boltzmann}
    p = p_\bullet[1+dT^\frac{1}{\alpha}]
\end{equation}
for $T\rightarrow0$.

\subsection{Andersen-Weeks-Chandler theory}

In their noted 1971 paper \cite{Andersen1971}, Andersen-Weeks-Chandler advises that the reference hard-sphere system should match the Helmholtz free energy of the potential in question. The effective hard-sphere diameter $d$ is determined within the theory from 
\begin{equation}\label{eq:AWC}
	\int_0^\infty y_\sigma(r)\Delta e(r)r^{d-1}dr = 0
\end{equation}
in which $\Delta e(r)=e(r)-e_\sigma(r)$ is the so-called blip function ($e(r)$ is defined in Eq.\ (\ref{eq:pair_boltzmann})), where and $y_\sigma(r)$ is the cavity function of the hard-sphere fluid. Within the Percus-Yevic approximation, the cavity function is given analytically \cite{Wertheim1963, Kahl1989, Chang1994, Hansen2013, Trokhymchuk2005, Smith2008, Henderson2009}. For the three-dimensional case ($d=3$), we used the implementation of the cavity function suggested by Chang et al. \cite{Chang1994}, as detailed in Section II.B.2 of Ref. \cite{Attia2022}. We evaluated the integral numerically using the Python module SciPy's \cite{SciPy} implementation of QUADPACK \cite{Piessens1983}.

At low temperatures, the blip function becomes narrow. Thus, it is appropriate to assume that the cavity function is constant. This approximation leads to the hard-sphere criterion due to Barker and Henderson \cite{Barker1967} (that predates the Andersen-Weeks-Chandler criterion). This allows for a closed form of the effective hard-sphere diameter for generalized Hertzian hyperspheres (see below).

\subsection{Low-temperature Barker-Henderson theory}\label{sec:Barker-Henderson}
Within Barker-Henderson theory \cite{Barker1967}, the effective hard-sphere diameter is given by
\begin{equation}\label{eq:bh}
    \sigma \equiv \int_0^\infty[1-e(r)]dr.
\end{equation}
In the low-temperature limit $(T\rightarrow0)$, the effective hard-sphere diameter simplifies to
\begin{eqnarray}\label{eq:barker-henderson_sigma}
    \sigma 
    &=& 1 - \Gamma\left(1+\frac{1}{\alpha}\right)T^{\frac{1}{\alpha}}.
\end{eqnarray}
and
\begin{equation}\label{eq:p_Barker-Henderson}
    p = p_\bullet\left[1+d\Gamma\left(1+\frac{1}{\alpha}\right)T^{\frac{1}{\alpha}}\right]\quad(T\rightarrow0).
\end{equation} 
for the coexistence pressure.

\subsection{The freezing line of Hertzian spheres}
To investigate the validity of the theory, we study Hertzian spheres ($d=3$; $\alpha=\frac{5}{2}$).
We use the \emph{interface pinning} \cite{Pedersen2013, Pedersen2013b, Thapar2014, Pedersen2015, Chew2023} combined with numerical integration of the Clausius-Clapeyron relation \cite{Kofke1993, Pedersen2023} to accurately determine the shape of the freezing line. 
The dashed line on Fig.\ \ref{fig:pressure} show the prediction from the naive hard-sphere mapping with $\sigma=1$ ($p_\bullet$; Eq.\ (\ref{eq:p_bullet_hs})). The coexistence pressure is significantly large within the hard-sphere paradigm, which is explained by the fact that the effective hard-sphere diameter is smaller. The normalized deviation from the naive prediction, $\frac{p}{p_\bullet}-1$, is depicted on a logarithmic scale in Fig.\ \ref{fig:pressure_log} as a solid black line. 
The dashed lines show the predictions due to the hard-sphere mappings. In agreement with Eqs.\ (\ref{eq:p_boltzmann}) and (\ref{eq:p_Barker-Henderson}) the deviation scales as $T^\frac{1}{\alpha}=T^\frac{2}{5}$. Interestingly, the involved Andersen-Weeks-Chandler theory (Eq.\ \ref{eq:AWC}) only gives marginally better prediction than the straightforward Barker-Henderson theory (Eq.\ (\ref{eq:p_Barker-Henderson})). In the low-temperature limit, the Barker-Henderson theory and Andersen-Weeks-Chandler theory provide a prediction within the statistical accuracy, while the Boltzmann prediction is slightly worse. Nonetheless, the Boltzmann criterion (Eq.\ (\ref{eq:p_boltzmann})) gives a good overall prediction.

\begin{figure}
    \centering
    \includegraphics[width=0.40\textwidth]{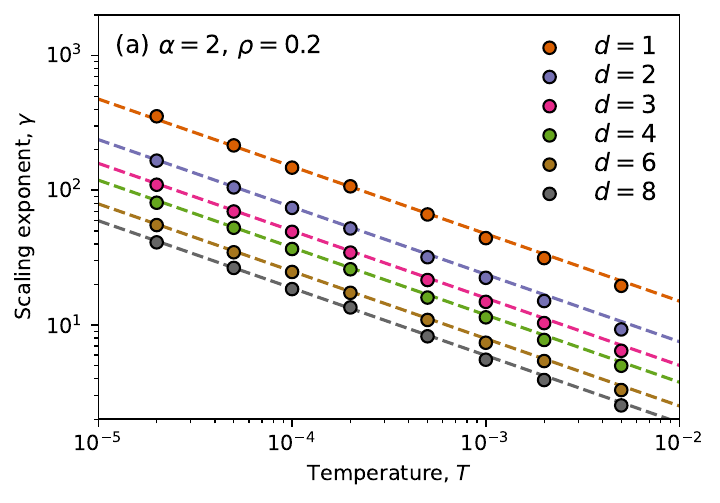}
    \includegraphics[width=0.40\textwidth]{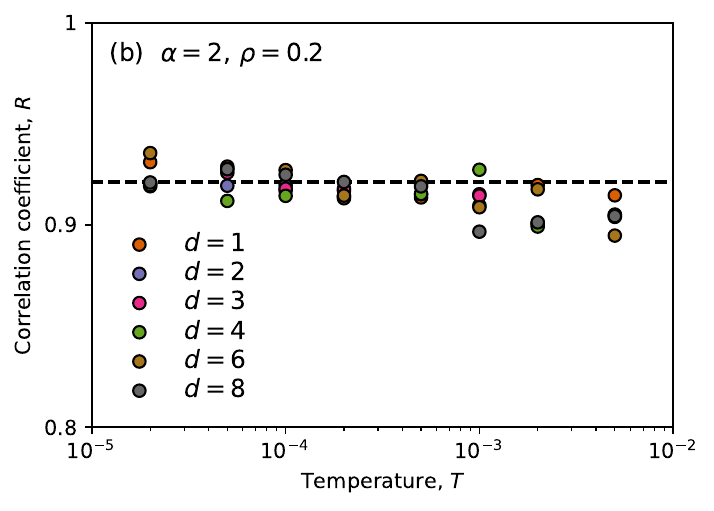}
    \includegraphics[width=0.40\textwidth]{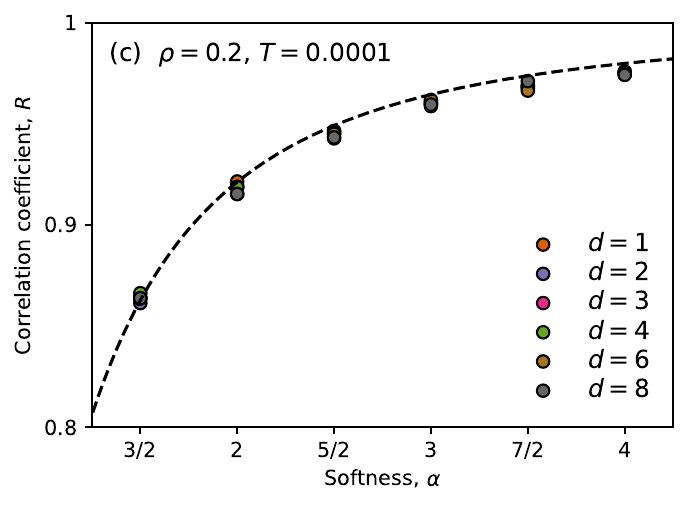}  
    \caption{(a) The dots show scaling exponent, $\gamma$ (Eq.\ (\ref{eq:gamma_definition})), computed in numerical simulation by evaluating Eq.\ (\ref{eq:gamma_nvt}) for hyperspheres with harmonic repulsions ($\alpha=2$) at density $\rho=0.2$ in $d=1$ to $d=8$ dimensions. The dashed lines are the theoretical predictions, Eq.\ (\ref{eq:gamma_theory}). (b) The dots show the virial-energy correlation coefficient, $R$ (Eq.\ (\ref{eq:R_definition})), evaluated in numerical simulations for a range of temperatures ($T$) and spatial dimensions ($d$). The dashed line is the theoretical prediction, Eq. (\ref{eq:R_theory}). (c) The dots show the virial-energy correlation coefficient, $R$ (Eq.\ (\ref{eq:R_definition})), along softness parameter, $\alpha$ for hyperspheres in different dimensions ($d$) at density $\rho=0.2$ and temperature $T=0.0001$.}
    \label{fig:gamma_R}
\end{figure}

\begin{figure}
    \centering
    \includegraphics[width=0.40\textwidth]{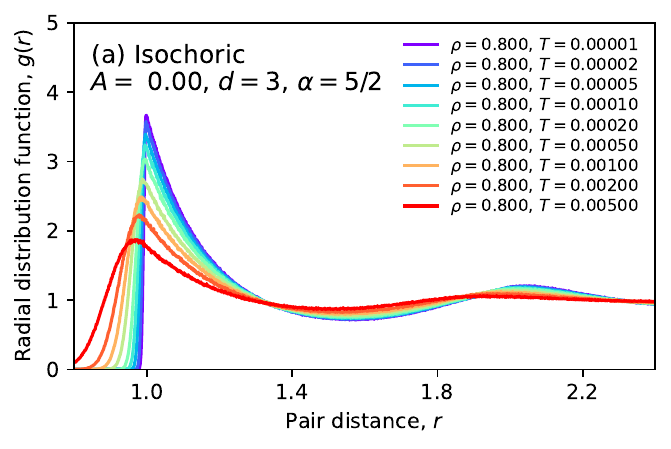}
    \includegraphics[width=0.40\textwidth]{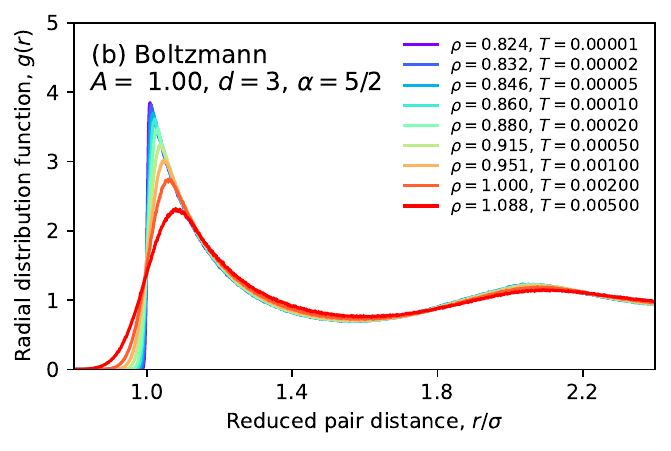}    
    \includegraphics[width=0.40\textwidth]{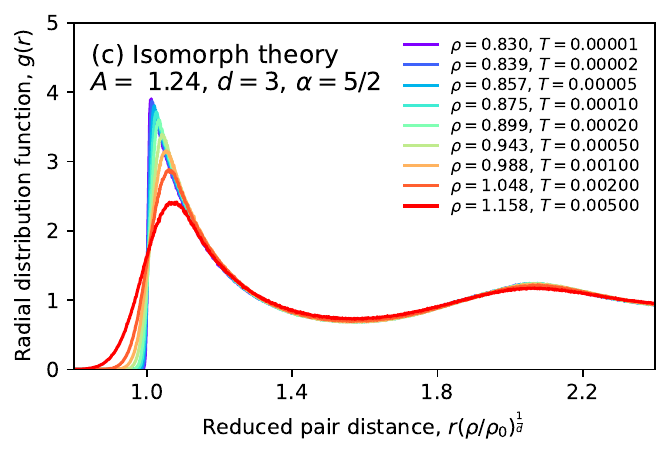}
    \includegraphics[width=0.40\textwidth]{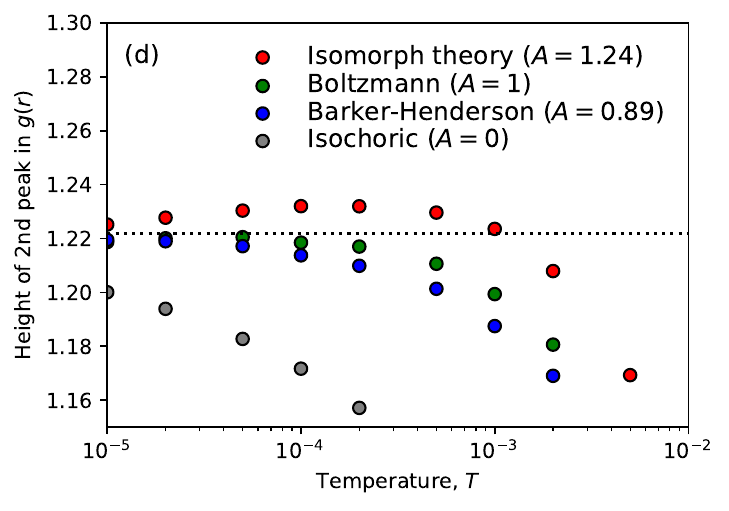}
    \caption{(a) The radial distribution, $g(r)$, of Hertzian spheres ($\alpha=\frac{5}{2}$; $d=3$) in the fluid state along the $\rho=0.8$ isochore. (b) The radial distribution, $g(r)$, of state points along the line of invariance using the Boltzmann criterion (Eq.\ (\ref{eq:scale_invariance_hs}) with $A=1$ and $\rho_0=0.8$) for the effective hard-sphere diameter ($\sigma$). The pair distances are in units of the effective hard-sphere diameter, $\sigma$. (c) The radial distribution, $g(r)$, along the isomorph suggested by Eq.\ (\ref{eq:scale_invariance_hs}) with $A=\Gamma(2+\frac{1}{\alpha})\simeq1.24$ and $\rho_0=0.8$, (see also Eq.\ (\ref{eq:rho_T_isomorph_theory})). The pair distances are scales by $(\rho_0/\rho)^\frac{1}{3}$. (d) The height of the second peak of $g(r)$ as a function of temperature. The dots are $\rho(T) $'s of the isomorph (red dots; see also panel (c)), the Boltzmann criterion (green; see also panel (b)), the Barker-Henderson hard-sphere mapping (blue), and the isomorph (gray; see also panel (a)). The dotted line indicates the low-temperature limit.}
    \label{fig:scaling_rdf}
\end{figure}

\begin{figure}
    \centering
    \includegraphics[width=0.4\textwidth]{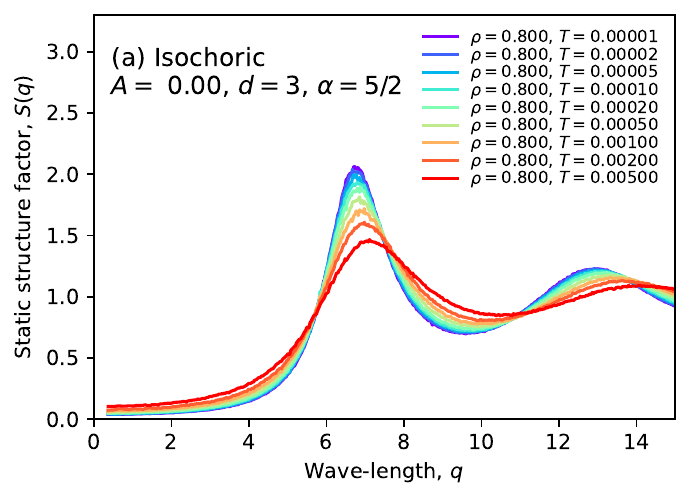}
    \includegraphics[width=0.4\textwidth]{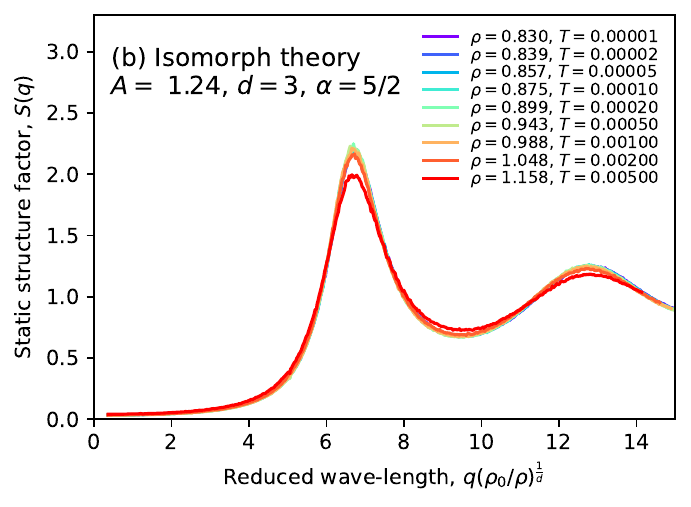}
    \includegraphics[width=0.4\textwidth]{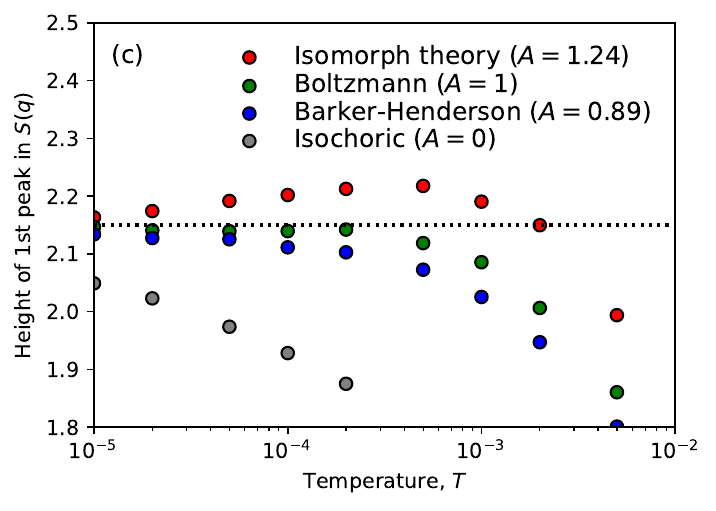}
    \caption{(a) The static structure factor, $S(q)$, f Hertzian spheres ($\alpha=\frac{5}{2}$; $d=3$) in the fluid state along the $\rho=0.8$ isochor. (b) The static structure factor, $S(q)$, along the isomorph suggested by Eq.\ (\ref{eq:scale_invariance_hs}) with $A=\Gamma(2+\frac{1}{\alpha})\simeq1.24$ and $\rho_0=0.8$. The length of the wave-vector, $q$, is scales by $(\rho/\rho_0)^\frac{1}{3}$. (c) The height of the first peak of the static structure factor along theoretical lines in the phase diagram with scale invariance.}
    \label{fig:scaling_Sq}
\end{figure}

\begin{figure}
    \centering
    \includegraphics[width=0.4\textwidth]{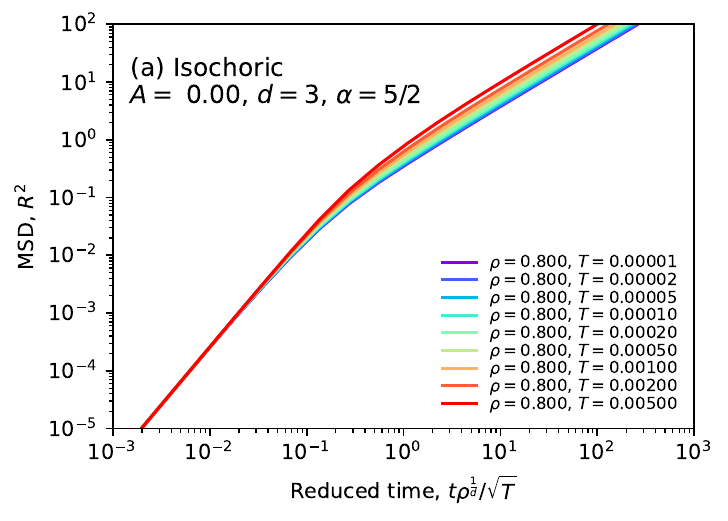}
    \includegraphics[width=0.4\textwidth]{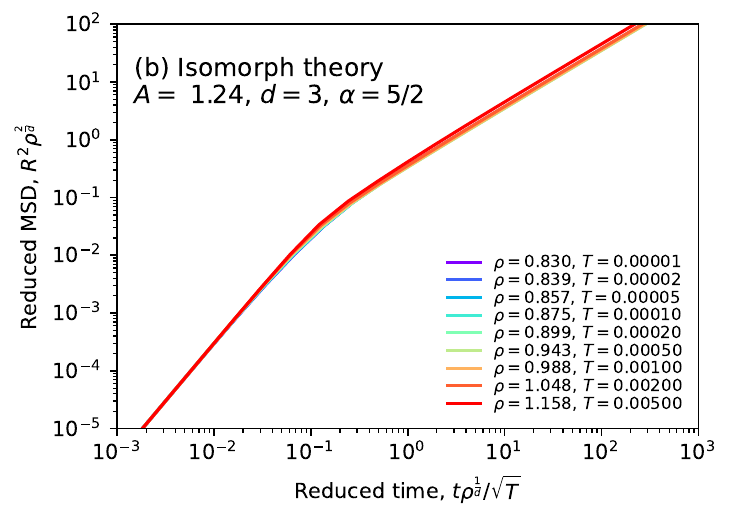}
    \includegraphics[width=0.4\textwidth]{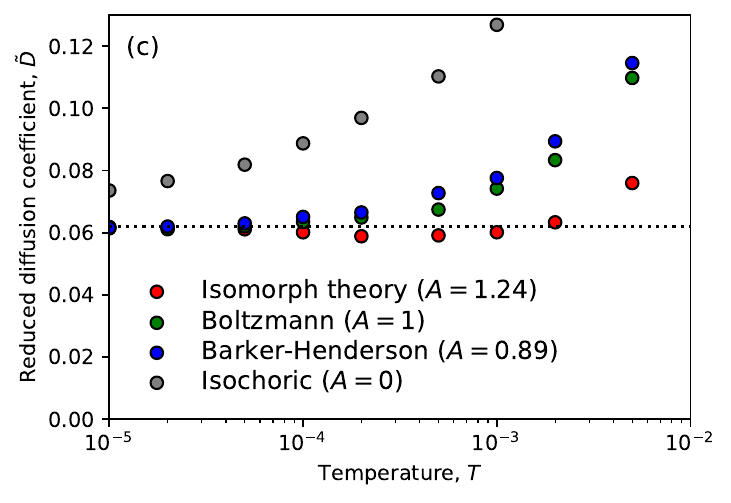}
    \caption{(a) The mean squared displace, $R^2$, of Hertzian spheres ($\alpha=\frac{5}{2}$; $d=3$) in the fluid state along the $\rho_0=0.8$ isochor as a function of reduced time, $\tilde t = t(\rho/\rho_0)^\frac{1}{d}/\sqrt{T}$. (b) The reduced mean squared displace, $\tilde R^2=R^2(\rho/\rho_0)^\frac{2}{3}$, as a function of reduced time ($\tilde t$) along the isomorph given by Eq.\ (\ref{eq:scale_invariance_hs}) with $A=\Gamma(2+\frac{1}{\alpha})\simeq1.24$ and $\rho_0=0.8$. (c) The reduced diffusion coefficient, $\tilde D$, as a function of temperature along theoretical lines of invariance.}
    \label{fig:scaling_msd}
\end{figure}

\section{Scale invariance}\label{sec:scaling_invariance}
The phase diagram of hard spheres is one-dimensional, as their structural and dynamical properties depend solely on the packing fraction, apart from a trivial temperature scaling. In contrast, the phase diagram of generalized Hertzian hyperspheres is two-dimensional, varying with density and temperature. Below, we present several frameworks for identifying lines in the phase diagram with scale invariance of structure and dynamics in reduced units. We present predictions based on mappings to the hard-hyperspheres \cite{Verlet1980, Verlet1981, Hansen2013, Dyre2016, Attia2022, Adhikari2023}, and apply the framework of isomorph theory \cite{Gnan2009, Dyre2014, Maimbourg2020, Attia2021}. Numerical simulations show that the theoretical predictions are excellent in providing lines of scale-invariance of structure and dynamics at sufficiently low temperatures.

\subsection{Hard spheres mapping}
As discussed in the previous sections \ref{sec:boltzmann} and \ref{sec:Barker-Henderson}, the effective hard-sphere diameter is
\begin{equation}
    \sigma = 1-AT^\frac{1}{\alpha}
\end{equation}
using Boltzmann's criterion for the diameter ($A=1$; Eq.\ (\ref{eq:boltzmann_sigma})) and Barker-Henderson theory ($A=\Gamma(1+\frac{1}{\alpha})$; Eq.\ (\ref{eq:barker-henderson_sigma})). For a state point at density $\rho$, the effective hard-sphere density is
\begin{eqnarray}
    \rho_\sigma &=& \rho\sigma^d \\
    &=& \rho [1-dAT^\frac{1}{\alpha}]\quad(T\to0).
\end{eqnarray}
Next, consider a reference state point $(\rho_0, T_0)$ and target state-point $(\rho, T)$. At low temperatures, we expect that generalized Hertzian hyperspheres approach the structure and dynamics of system hard hyperspheres. Let \emph{hard-sphere units} be a system where $\sigma$ and $T$ are unity, e.g., a length is reported as $\tilde d=d/\sigma$. Invariant structure and dynamics in hard-sphere units are expected for the two state-points if
\begin{equation}
    \rho [1-dAT^\frac{1}{\alpha}] = \rho_\star [1-dAT_\star^\frac{1}{\alpha}].
\end{equation}
Assuming that both $T\ll1$ and $T_\star\ll0$, then the temperature dependence of state-points with the same structure and dynamics is given by
\begin{equation}\label{eq:scale_invariance_hs}
    \rho(T) = \rho_0[1+dAT^\frac{1}{\alpha}]\quad(T\to0)
\end{equation}
where 
\begin{equation}
\rho_0=\rho_\star[1-dAT_\star^\frac{1}{\alpha}] 
\end{equation}
is a constant.

\subsection{Isomorph theory}

The fundamental assumption of isomorph theory \cite{Pedersen2008, Gnan2009, Schroeder2014, Pedersen2016} is that the potential energy possesses \emph{hidden scale invariance}: Consider two physically relevant configurations, ${\bf R}_\mathcal{A}$ and ${\bf R}_\mathcal{B}$, typical in the canonical ensemble (at a given state point). Let $\zeta$ be an affine-scaling parameter. The energy function exhibits hidden scale invariance if an affine scaling preserves the rank order of energies, meaning that if \cite{Schroeder2014}
\begin{equation}\label{eq:U_scale_invariance}
    U({\bf R}_\mathcal{A}) < U({\bf R}_\mathcal{B}) \Rightarrow 
    U(\zeta{\bf R}_\mathcal{A}) < U(\zeta{\bf R}_\mathcal{B})
\end{equation}
holds to a good approximation. From this, it follows that the Person correlation coefficient between viral and energy,
\begin{equation}\label{eq:R_definition}
    R = \frac{\langle wv\rangle-\langle w\rangle\langle v\rangle}{\sqrt{(\langle w^2\rangle-\langle w\rangle^2)(\langle v^2\rangle-\langle v\rangle^2)}},
\end{equation}
is close to unity. The correlation coefficient is a practical measure of the degree to which the system possesses hidden scale invariance \cite{Pedersen2008, Gnan2009}.
Moreover, isomorph theory predicts that structure and dynamics in units reduced by $\rho$ and $T$ are invariant along the lines of constant excess entropy, $S_\textrm{ex}$, also referred to as \emph{Rosenfeld's entropy scaling} \cite{Rosenfeld1977, Rosenfeld1999, Pond2011, Dyre2018}. 
The first step when applying isomorph theory is to identify lines of constant excess entropy with the slope
\begin{equation}\label{eq:gamma_definition}
    \gamma\equiv\left.\frac{\partial\ln T}{\partial\ln \rho}\right|_{S_\textrm{ex}}
\end{equation}
From basic thermodynamic identities, it follows that
\begin{equation}\label{eq:gamma_nvt}
    \gamma = \frac{\langle wv\rangle-\langle w\rangle\langle v\rangle}{\langle v^2\rangle-\langle v\rangle^2}.
\end{equation}
Following the theoretical insights presented in Section \ref{sec:theory}, we consider the low-density limit \cite{Maimbourg2016, Bacher2018, Bacher2018b, Maimbourg2020, Attia2021}, $\rho\to0$, 
where $Z_s\rightarrow0$. We can then neglect terms that involve multiplications of expectation values since terms that involve a single expectation value scale as $1/Z_s$, while terms involving multiplication of expectation values scale as $1/Z_s^2$. Thus, the scaling exponent and correlation coefficient are simplified to
$
\gamma = \langle wv\rangle/\langle v^2\rangle
$
and
$
R = \langle wv\rangle/\sqrt{\langle w^2\rangle\langle v^2\rangle},
$
respectively.
From Eq.\ (\ref{eq:w_iu_i}) this can be written as
$\gamma = I_{d\alpha11}/I_{d\alpha02}
$
and
$
R = I_{d\alpha11}/\sqrt{I_{d\alpha20} I_{d\alpha02}}.
$
By insertion of Eq.\ (\ref{eq:I_daij}) we arrive at
\begin{equation}\label{eq:gamma_theory}
    \gamma = \frac{\alpha/d}{\Gamma(2+\frac{1}{\alpha})}T^{-\frac{1}{\alpha}} \quad(T\to0).
\end{equation}
The scaling exponent diverges at $T\to0$. This \emph{extreme density scaling} is also observed for the noted Weeks-Chandler-Andersen pair potential \cite{Attia2021}.
For the correlation coefficient, the prediction for the low-temperature value saturates at
\begin{equation}\label{eq:R_theory}
    R = \frac{1}{\sqrt{\Gamma(2+\frac{1}{\alpha})\Gamma(2-\frac{1}{\alpha}))}}\quad(T\to0).
\end{equation}
Interestingly, the correlation coefficient approached a value less than one for $T\to0$. This value is independent of spatial dimension ($d$) but depends on the softness of the pair potential ($\alpha$). In limit $\alpha\to\infty$ the correlation coefficient approaches one ($\Gamma(2)=1$).

By integration of Eq.\ (\ref{eq:gamma_theory}), and assuming that $T\ll1$, we can find the shape of the lines with constant excess entropy ($S_\textrm{ex}$), i.e., the isomorph:
\begin{equation}\label{eq:rho_T_isomorph_theory}
    \rho(T) = \rho_0\left[1+d\Gamma\left(2+\frac{1}{\alpha}\right)T^{\frac{1}{\alpha}})\right].
\end{equation}
where $\rho_0$ is a constant.
This is identical to Eq.\ (\ref{eq:scale_invariance_hs}), derived by hard-sphere mapping, if $A=\Gamma(2+\frac{1}{\alpha})$. Thus, the theoretical predictions using hard-sphere mapping or isomorph theory predict the same shape of the lines with scale-invariant dynamics and structure ($T^\frac{1}{\alpha}$ temperature dependence of the density) but with different $A$ coefficients.

\subsection{Comparing scaling laws}
Above, we derive several theoretical predictions for lines in the $(\rho, T)$ phase diagram with invariant dynamics and structure. In the following, we compare the theories and asses their applicability by numerical simulations. 

Figure \ref{fig:scaling_rdf}(a) shows how the radial distribution function, $g(r)$, for Hertzian spheres ($d=3$; $\alpha=\frac{5}{2}$) at density $\rho=0.8$ changes in the temperature interval $T=10^{-5}$ to $T=5\times10^{-3}$. For the same temperature range, Fig.\ \ref{fig:scaling_rdf}(b) shows $g(r)$ for state points given by Eq.\ (\ref{eq:scale_invariance_hs}) with $A=1$ and $\rho_0=0.8$, where the pair distances have been scaled by the effective hard-sphere diameter ($\sigma$). Interestingly, with the exception of the first peak, which is influenced by the shape of the pair potential, the remainder of this structural behavior is invariant (compared to the isochoric state points in Fig.\ \ref{fig:scaling_rdf}(a)). 
Figure \ref{fig:scaling_rdf}(c) shows the isomorph prediction of Eq.\ (\ref{eq:scale_invariance_hs}) with $A=\Gamma(2+\frac{1}{\alpha})\simeq1.24$ and $\rho_0=0.8$, giving scale invariance similar to that of Boltzmann's hard-sphere mapping. To quantify the scaling laws, Fig.\ \ref{fig:scaling_rdf}(d) shows the height of the second peak of $g(r)$. Surprisingly, Boltzmann's criterion (green dots) predicts better than Barker-Henderson's (blue dots), while isomorph theory provides the best overall scale invariance for the investigated temperature range. 
Investigating the static structure factor, $S(q)$, can be instructive (see Figs.\ \ref{fig:scaling_Sq}(a)-\ref{fig:scaling_Sq}(c)) to avoid the trivial non-scaling of the first peak of the radial distribution function ($g(r)$). Figures \ref{fig:scaling_msd}(a)-\ref{fig:scaling_msd}(c) show scale invariance of the mean squared displacement ($R^2$) and the diffusion coefficient. For the dynamics, the isomorph theory gives the overall best scale invariance.

\begin{acknowledgments}
The author is grateful for the discussions with Lorenzo Costigliola, Jeppe C. Dyre, and David Hayes. This work was supported by the VILLUM Foundation's Matter Grant (No. 16515).
\end{acknowledgments}

\section*{Data Availability}
The data that support the findings of this study are openly available in Zenodo at \href{http://doi.org/10.5281/zenodo.14850597}{http://doi.org/10.5281/zenodo.14850597}, reference number 10.5281/zenodo.14850597.

\appendix
\section{Solution to Eq.\ (\ref{eq:Idaij_definition})}
\label{appendix:deviation}
We solve the integral of Eq.\ (\ref{eq:Idaij_definition}) by substituting $t=T^{-1}(1-r)^\alpha$ as integration variable so $1-r=(tT)^\frac{1}{\alpha}$, $r=1-(tT)^\frac{1}{\alpha}$ and $dr = -\alpha^{-1}T^\frac{1}{\alpha}t^{\frac{1}{\alpha}-1}dt$.
In the $T\to0$ limit the $I_{d\alpha ij}$ simplifies to an expression involving the $\Gamma$-function (Eq.\ (\ref{eq:gamma_function})):
\begin{widetext}
\begin{eqnarray}
    I_{d\alpha ij} &=& \int_0^1 [w(r)]^i[v(r)]^j e(r) d{\bf r} \\
    &=& \int_0^1 \left[\frac{\alpha}{d}r(1-r)^{\alpha-1}\right]^i[(1-r)^\alpha]^j \exp(-T^{-1}(1-r)^\alpha) S_dr^{d-1}dr\\
    &=& \frac{S_d\alpha^i}{d^i}\int_0^1 r^{i+d-1} [1-r]^{j\alpha+i[\alpha-1]}\exp(-T^{-1}(1-r)^\alpha)dr\\
    &=& \frac{S_d\alpha^i}{d^i}\int_{T^{-1}}^0 [1-(tT)^\frac{1}{\alpha}]^{i+d-1} [(tT)^\frac{1}{\alpha}]^{j\alpha+i[\alpha-1]}\exp(-t)[-\alpha^{-1}T^\frac{1}{\alpha}t^{\frac{1}{\alpha}-1}]dt \quad (t=T^{-1}(1-r)^\alpha) \\
    &=& \frac{S_d\alpha^{i-1}T^{i[1-\frac{1}{\alpha}]+j+\frac{1}{\alpha}}}{d^i}\int^{T^{-1}}_0 t^{i[1-\frac{1}{\alpha}]+j+\frac{1}{\alpha}-1} [1-(tT)^\frac{1}{\alpha}]^{d+i-1}\exp(-t)dt  \\
    &\to& \frac{S_d\alpha^{i-1}T^{i[1-\frac{1}{\alpha}]+j+\frac{1}{\alpha}}}{d^i}\int^\infty_0 t^{i[1-\frac{1}{\alpha}]+j+\frac{1}{\alpha}-1} \exp(-t)dt \quad (T\to0)  \\
    &=& \frac{S_d\alpha^{i-1}T^{i+j+\frac{1-i}{\alpha}}}{d^i}\Gamma\left(i+j+\frac{1-i}{\alpha}\right)  \\
    &=& \frac{S_d\alpha^{i-1}T^{J_{\alpha ij}}}{d^i}\Gamma\left(J_{\alpha ij}\right)
\end{eqnarray}
where $J_{\alpha ij}\equiv i+j+\frac{1-i}{\alpha}$. 
\end{widetext}

\bibliography{references}

\end{document}